%% file: ofj-template.tex
\RequirePackage[hyphens]{url}

\documentclass[e-only,10pt,reqno]{ofj}

\usepackage{setspace}
\usepackage{lineno}
\usepackage{color}
\usepackage[symbol]{footmisc}
\usepackage[dvipsnames,svgnames,x11names]{xcolor}
\usepackage{siunitx}
\usepackage{orcidlink}

\newcommand{\authorcontributions}[1]{%
\vspace{6pt}\noindent{\fontsize{9}{11.2}\selectfont\textbf{Author Contributions:} {#1}\par}}

\let\paragraph\undefined

\copyrightinfo{2022}{The Authors. This work is an open access article under the CC BY-SA 4.0 license}

\newcommand{\OF}[0]{OpenFOAM\textsuperscript{\textregistered} }

\OpenFOAMversions{\OF v2106}
\Repository{https://github.com/simzero-oss/rom-js}

\DOI{10.51560/ofj.v2.83} 

\begin{document}


\title[\hfill\protect\parbox{0.975\linewidth}{rom.js/cfd.xyz: An open-source framework for generating and \protect\\ visualizing parametric CFD results}]{rom.js/cfd.xyz: An open-source framework for generating and visualizing parametric CFD results}




\author{C. Pe\~{n}a-Monferrer$^{*}$\orcidlink{0000-0003-3271-6399}}
\address{SIMZERO, Spain}
\email{carlos@simzero.com}

\author{C. Díaz-Marín\orcidlink{0000-0002-8924-9544}}


\begin{abstract}
We present in this technical note an open-source web framework for the generation and visualization of parametric OpenFOAM simulations from surrogate models. It consists of a JavaScript module (rom.js) and a web app (cfd.xyz) to explore fluid dynamics problems efficiently and easily for a wide range of parameters. rom.js is a JavaScript port of a set of open-source packages (Eigen, Splinter, VTK/C++ and ITHACA-FV) to solve the online stage of reduced-order models (ROM) generated by the ITHACA-FV tool. It can be executed outside a web browser within a backend JavaScript runtime environment, or in a given web solution. This methodology can also be extended to methods using machine learning. The rom.js module was used in cfd.xyz, an open-source web service to deliver a collection of interactive CFD cases in a parametric space. The framework includes some tutorials, showing the whole process from the generation of the surrogate model to the web browser. It also includes a standalone web tool for visualizing users' ROMs by directly dragging and dropping the output folder of the offline stage. Beyond the current proof of technology, this enables a collaborative effort for the implementation of OpenFOAM surrogate models in applications demanding real-time solutions such as digital twins and other digital transformation technologies.
\end{abstract}

\date{\today}

\dedicatory{}

\maketitle



\section{Introduction}

Recent technological developments have made it possible to accelerate computational fluid dynamics (CFD) modeling through physics-based or data-driven surrogate modeling. This acceleration is key for enabling the integration of such solutions with real-life systems for dynamically controlling complex processes, but also allows an interactive analysis of the parametric space of CFD simulations.

Different techniques using machine learning (ML) and/or reduced-order modeling (ROM) applied to CFD have been described in the literature \cite{GHNATIOS201229,DING2019106394,GANTI2020104626,Stabile2017CAIM, Stabile2017CAF,Kochkove2101784118}. Furthermore, several open-source packages have been released in the past few years \cite{AccelerateCFD, ITHACA-FV, AndreWeiner, PranshuPant, PythonFOAM}. We aim with the developments presented in this work to create a shared space where canonical and industrial CFD problems can be visualized and analyzed without carrying out a simulation, or as a preliminary step for optimizing parameters of new simulations. Having an open-source centralized service has several advantages, not only from an educational, optimization and reproducibility point of views but also from a CO\textsubscript{2} footprint perspective. Computation and data processing are associated with increasing greenhouse gas emissions \cite{zwart_ecological_2020,freitag2021real,strubell_energy_2019} that are expected to increase significantly in the coming years with the use of high-performance computing simulations and ML.

With our development, we predict the following benefits for the community:

\begin{itemize}
    \item Educational enhancement: The use of the web app can accelerate the learning of fluid dynamics by reducing the entry barrier of simulations. For example, a better understanding of how viscosity, gravity or inlet velocity affects fluid dynamics could be easily observed directly in the browser without any software installation or previous knowledge of how to use that software.
    \item Optimization: A preliminary analysis of similar physical problems through data-driven solutions can provide useful information for a given industrial application. This might help at early stage of designs by optimizing time and resources.
    \item Reproducibility: The framework provides reproducible examples of CFD and ROM for OpenFOAM tutorials that can be extended to cover more cases.
    \item Worldwide duplicity reduction: The use of CFD simulations around the world is increasing rapidly and is applied nowadays from very specific industrial fields to daily applications. While this is undoubtedly deepening our knowledge of fluid mechanics in industrial applications, it results in an inefficient use of resources in global terms. Full-order model simulations will still be needed for many research and industrial activities, however the adoption of a collaborative open-source web framework for surrogate models could save many executions of simulations of similar scenarios, thus reducing the carbon footprint.
    \item CFD sharing flexibility: The framework was designed to welcome open-source contributions of CFD cases (offline stage), but also to only showcase the surrogate model, thereby being compatible with organizations willing to protect their CFD models.
\end{itemize}

In addition, the web app also provides a standalone tool that directly connects the files of the surrogate model with the parametric visualization by only dragging and dropping the ROM output folder. This results in a convenient way of inspecting surrogate models, as you can directly visualize the results in-situ without generating a data file per parameter.

Different open-source tools for scientific visualization are already available. Kitware \cite{kitwarewebsite} developed a variety of software such as vtk.js \cite{vtkjsWeb}, ParaViewWeb \cite{paraviewWeb}, ParaView Glance \cite{paraviewGlanceWeb} and trame \cite{trame} for visualizing scientific data on the web. Regarding the visualization of parametric simulations, RBniCS \cite{HesthavenRozzaStamm2015,RBniCS} developed a Python framework for reduced-order modeling with FEniCS. Examples using RBniCS are shown in ARGOS \cite{ARGOS} for different mathematical problems. Finally, ParaView recently integrated a Python plugin for viewing inference results and monitoring the training process in real time for deep-learning surrogates \cite{kitwareDeepLearning}. Compared with these frameworks, our approach provides: a) an integrated solution for OpenFOAM from the parallel execution of simulations natively to ROM visualization, b) a JavaScript port of the required tools for solving both turbulent ROM online stage and post-processing results for visualization, c) an integrated web app for interactively visualizing the results for the different parameters.

We developed two open-source tool tools for this work, rom.js \cite{rom.js} and cfd.xyz \cite{cfd.xyz}. rom.js is a JavaScript port of different packages such as Eigen \cite{eigenweb}, Splinter \cite{splinter}, VTK/C++ \cite{vtk}, and ITHACA-FV \cite{Stabile2017CAIM, Stabile2017CAF, ITHACA-FV} for interactively solving the online stage of reduced-order models generated by ITHACA-FV. Finally, cfd.xyz is a web app that integrates rom.js and vtk.js to deliver a user-friendly, modular tool. A starting point for the project is to showcase an example for a turbulent steady-state OpenFOAM tutorial (pitzDaily) showing how to generate the surrogate models and integrate them in the web app. Future work will focus on integrating other ROM techniques, ML and CFD packages on the framework and incorporating other industrial problems.

\section{Framework description}

Our work relies on the offline-online splitting technique for facilitating the synergy between high performance computing and reduced order methods \cite{salmoiraghi2016advances}. In the offline stage, we compute on an HPC or a workstation the full-order model (FOM) for a selected set of parameters. In the online stage, we evaluate the solution with new parameters at significantly lower computational cost.

We created a JavaScript module, rom.js, containing the code needed for solving the online stage of reduce-order models. We also developed some other tools for facilitating the execution of the  offline stage runs and the generation of the ROM. The module can be directly imported to work alongside JavaScript on a web app or in a backend runtime. To showcase its use we created cfd.xyz, an open-source and cross-platform web app for generating and visualizing CFD data. The results shown in this work can be reproduced with the v1.0.0-rc.8 versions of these repositories.

\subsection{rom.js - An open-source JavaScript module for the online stage of reduced-order modeling}

The rom.js module is a new implementation with ported versions of Eigen, Splinter, VTK/C++ and ITHACA-FV to compute and visualize the online stage of ROM generated by ITHACA-FV. They were ported to WebAssembly \cite{RossbergWebAssemblyCoreSpecification} using Emscripten. WebAssembly is a portable binary-code format that provides a way to run code written in different languages on the web at near-native performance. Emscripten is a compiler toolchain that compiles the source code to WebAssembly. The module resulted in an optimized single JavaScript file of around 7 MB, which is a relatively small size considering that it can itself compute the ROMs solution and perform the volumetric post-processing tasks needed for visualization. 

The main use of Eigen in this package is dealing with matrix operations and solving the ROM online stage. Only a small subset of Eigen was needed. Splinter was ported to compute the online turbulent viscosity using radial basis functions as described in \cite{hijazi_data-driven_2020}. Finally, VTK/C++ was used to reconstruct the calculated values from Eigen to the original VTK unstructured grid, and to apply VTK filters. It is worth noticing that vtk.js supports web visualization, but not unstructured grids and related components. The integration of VTK/C++ in rom.js allows the reconstruction of the calculated volumetric fields to VTK format. It also enables a later connection with vtk.js components for the visualization in a web browser environment.

\begin{figure}{!ht}
\includegraphics[width=1.0\textwidth]{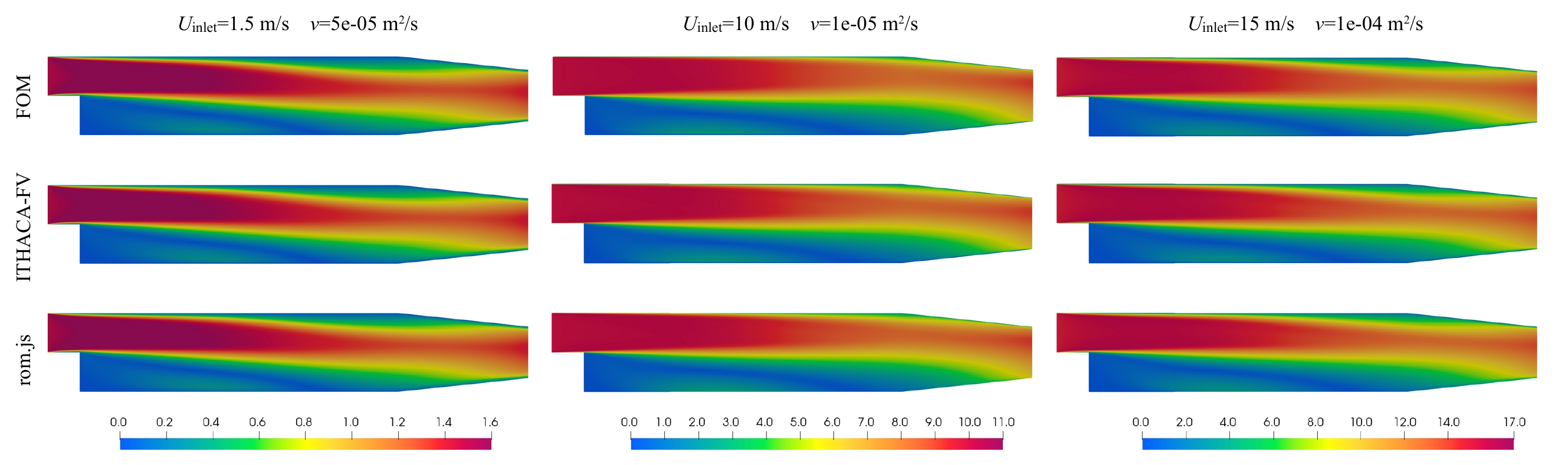}
\caption{Comparison of velocity magnitude for OpenFOAM full-order model (FOM), ITHACA-FV (ROM) and rom.js (ROM) for $U_{\textrm{inlet}}$ of 1.5, 10.0 and \SI{15,0}{\meter/\second} and kinematic viscosity, $\nu$, of \num{5e-05}, \num{1e-05} and \SI{1e-04}{\meter\squared/\second}.}
\label{fig:verification}
\end{figure}

A workflow was developed for generating the CFD snapshots directly with OpenFOAM for the offline stage. In this stage, the full-order model was solved repeatedly to construct the parametric space by varying inlet velocity and kinematic viscosity in the range of 0.5 to 20 m/s and \num{5e-06} to \SI{1e-04}{\meter\squared/\second}, respectively. These snapshots were later used for building the ROM with a new application based on the ITHACA-FV library. Around 800 simulations were performed using a massively parallel processing approach of single core runs for the pitzDaily tutorial. Within this approach, a bash script was designed to set each simulation with the corresponding parameters and execute batches of simulations. The size of the batches is defined as a user input with the total number of cores to allocate for this task. The turbulent intensity is fixed at the inlet to 5\% for this example, and therefore values for turbulent kinetic energy ($\kappa$) and its dissipation ($\varepsilon$) are recalculated according to every inlet velocity.

The current version supports turbulent and laminar, incompressible, steady-state cases. For the turbulent case, the ROM is generated using a mixed strategy that combines a data-driven reduction method to approximate the eddy viscosity solution manifold and a classical POD-Galerkin projection approach for velocity and pressure fields (see \cite{hijazi_data-driven_2020} for further details). The stabilization strategy used in our work is based on the supremizer enrichment \cite{Stabile2017CAF}. The ROM was built with 16 modes for velocity, and 5 modes for pressure, eddy viscosity and supremizers.

The ROM online stage can be then solved with the provided velocity inlet and viscosity user inputs. The total execution time for calculating the new fields and reconstructing them with rom.js was around 0.027 s. on an AMD Ryzen 9 5950X @ 3.4 GHz with 64GB of RAM. A comparison of the velocity magnitude for OpenFOAM (FOM), ITHACA-FV C++ (ROM) and the rom.js (ROM) ported version are shown in Fig.~\ref{fig:verification} for three conditions. This figure shows a good visual agreement between the FOM and ROM for the three different conditions. It also verifies that the rom.js implementation gives good results compared with ITHACA-FV. A mean root-mean-square error of \num{6.2e-04}, \num{6.4e-04} and \num{1.4e-04} m/s was obtained for the comparison between ITHACA-FV and rom.js for an $U_{\textrm{inlet}}$ of 1.5, 10.0 and 15.0 m/s, respectively. Regarding the comparison between the OpenFOAM's FOM and rom.js, the root-mean-square error resulted in 0.0133, 0.0381 and 0.0127 m/s, respectively.

\subsection{cfd.xyz - An open-source web platform for generating and visualizing of CFD data}

\begin{figure}
\includegraphics[width=1.0\textwidth]{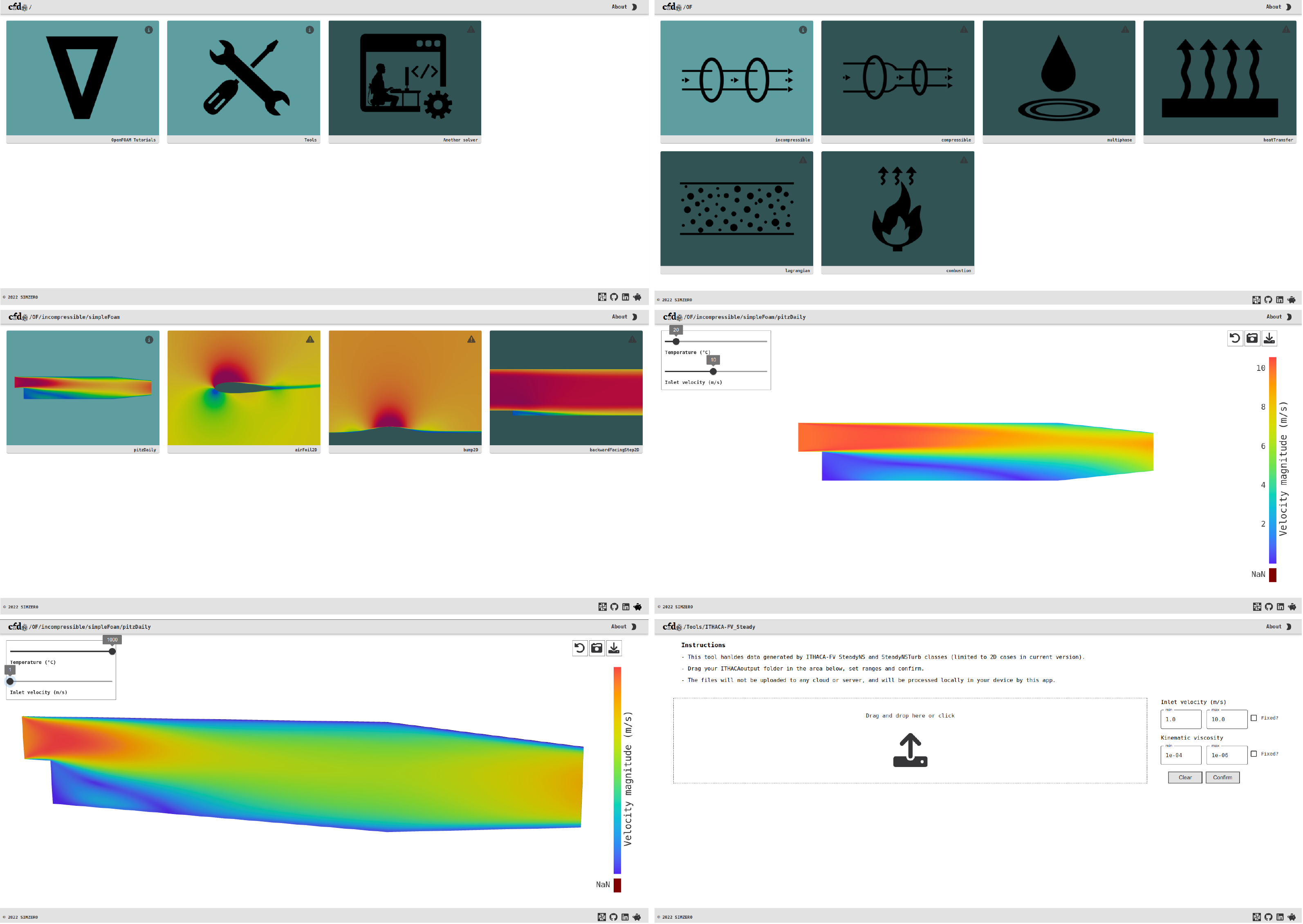}
\caption{Screenshots of cfd.xyz for different pages.}
\label{fig:screenshots}
\end{figure}

A single page web app, cfd.xyz, was designed in a modular way to include content containers in the form of cards for the different OpenFOAM tutorials. The ROM generated by ITHAFCA-FV was used by the web app as the surrogate models. Every tutorial available in the platform contains the interactive visualization of a case for a range of different parameters (e.g., viscosity, inlet velocity, ...). The data for a given set of parameters can also be downloaded for further post-processing on a dedicated desktop or web tool. Fig.~\ref{fig:screenshots} shows screenshots of different stages of the web app.

\begin{figure}
\includegraphics[width=0.8\textwidth]{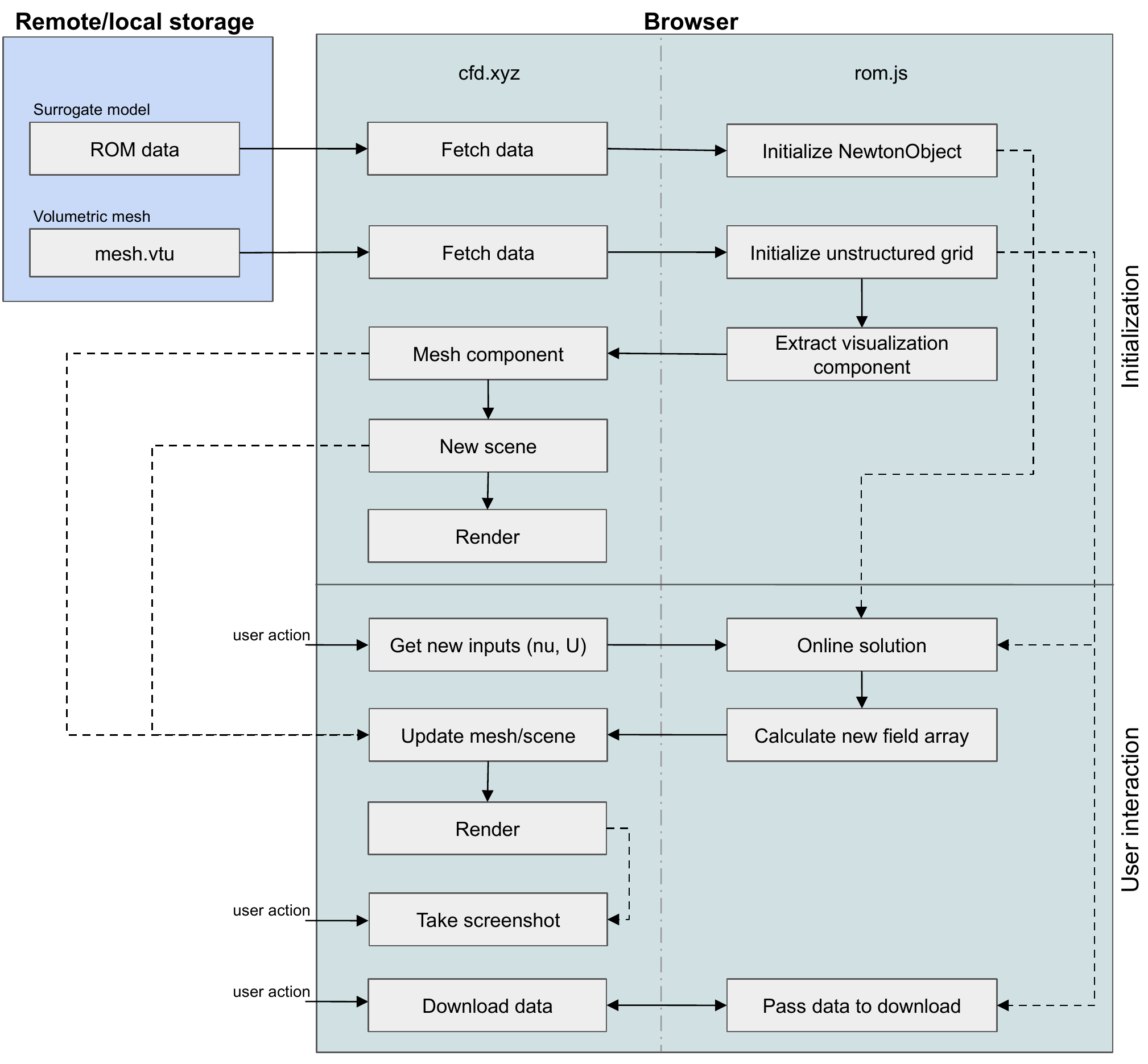}
\caption{High-level overview of the components in cfd.xyz }
\label{fig:overview}
\end{figure}

The web app was built with React JS, a JavaScript library for building user interfaces. It is client-based and the communication with the server is mainly for sending the bundle to the user. The visualization of the cases for every set of parameters was achieved with vtk.js and rom.js. Fig.~\ref{fig:overview} shows an overview of this process for a given case. The volumetric mesh (unstructured grid) and ROM data are fetched from remote or local storage and initialized in the WebAssembly side with rom.js. A rendering scene is initially set-up and rendered. The user can interact with the 3D view and modify the parameters defined for the case. A change of these parameters triggers the calculation and visualization of the new fields. Additionally, the data can be directly visualized or downloaded as VTK files or images.

\section{Conclusion}

We presented in this technical note an open-source framework for parametric generation and visualization of CFD data.  This approach provides an integrated solution for OpenFOAM cases from the parallel execution of simulations natively with OpenFOAM to the visualization of results from reduced-order models. A JavaScript port was performed for the required tools for solving turbulent ROM online stage and post-processing results for visualization. Finally, a web app was developed for interactively visualizing the results for the different parameters.

The framework consisted of two main components: a JavaScript module (rom.js) and a React JS web app (cfd.xyz). Working together, these tools create a shared space where canonical and industrial CFD problems can be visualized and analyzed on the web without carrying out a simulation. The models and workflows for generating the surrogate models are also shared in a reproducible way. New cases can be easily integrated on the web with the implemented methodology. The current version is based on OpenFOAM as the CFD tool and ITHACA-FV as the reduced-order modeling (ROM) one. Further implementations will be considered for including other ROM or machine learning packages, as well as other CFD tools.

The pitzDaily OpenFOAM tutorial was used as an example. The generated ROM showed a good agreement compared with the full-order model. The performance of the rom.js ported implementation was also satisfactorily verified. This module resulted in a JavaScript file of 7 MB and can itself compute the ROMs solution and perform the volumetric post-processing tasks needed for visualization. The calculation and reconstruction of the new fields for a given set of parameters resulted in an execution time of around 0.027 s. using this module. This permitted a smooth interaction through the different parameters on the web version, and enables future implementations of new tools using CFD.


%



\authorcontributions{
Conceptualisation, C.P-M. and C.D-M.;
methodology, C.P-M.;
software, C.P-M;
validation, C.P-M.;
formal analysis, C.P-M.;
investigation, C.P-M.;
resources, C.P-M.;
data curation, C.P-M. and C.D-M.;
writing---original draft preparation, C.P-M. and C.D-M;
writing---review and editing, C.P-M.;
visualisation, C.P-M. and C.D-M;
supervision, C.P-M.;
project administration, C.P-M.;
All authors have read and agreed to the published version of the manuscript.
}


%
%

\bibliographystyle{IEEEtran}

\input{ofj-template.bbl}


\end{document}

%% file: ofj-template.bbl

%% file: ofj-template.bbl
\begin{thebibliography}{10}
\providecommand{\url}[1]{#1}
\csname url@samestyle\endcsname
\providecommand{\newblock}{\relax}
\providecommand{\bibinfo}[2]{#2}
\providecommand{\BIBentrySTDinterwordspacing}{\spaceskip=0pt\relax}
\providecommand{\BIBentryALTinterwordstretchfactor}{4}
\providecommand{\BIBentryALTinterwordspacing}{\spaceskip=\fontdimen2\font plus
\BIBentryALTinterwordstretchfactor\fontdimen3\font minus
  \fontdimen4\font\relax}
\providecommand{\BIBforeignlanguage}[2]{{%
\expandafter\ifx\csname l@#1\endcsname\relax
\typeout{** WARNING: IEEEtran.bst: No hyphenation pattern has been}%
\typeout{** loaded for the language `#1'. Using the pattern for}%
\typeout{** the default language instead.}%
\else
\language=\csname l@#1\endcsname
\fi
#2}}
\providecommand{\BIBdecl}{\relax}
\BIBdecl

\bibitem{GHNATIOS201229}
\BIBentryALTinterwordspacing
C.~Ghnatios, F.~Masson, A.~Huerta, A.~Leygue, E.~Cueto, and F.~Chinesta,
  ``Proper generalized decomposition based dynamic data-driven control of
  thermal processes,'' \emph{Computer Methods in Applied Mechanics and
  Engineering}, vol. 213-216, pp. 29--41, 2012. [Online]. Available:
  \url{https://www.sciencedirect.com/science/article/pii/S0045782511003641}
\BIBentrySTDinterwordspacing

\bibitem{DING2019106394}
\BIBentryALTinterwordspacing
C.~Ding and K.~P. Lam, ``Data-driven model for cross ventilation potential in
  high-density cities based on coupled {CFD} simulation and machine learning,''
  \emph{Building and Environment}, vol. 165, p. 106394, 2019. [Online].
  Available:
  \url{https://www.sciencedirect.com/science/article/pii/S0360132319306043}
\BIBentrySTDinterwordspacing

\bibitem{GANTI2020104626}
\BIBentryALTinterwordspacing
H.~Ganti and P.~Khare, ``Data-driven surrogate modeling of multiphase flows
  using machine learning techniques,'' \emph{Computers \& Fluids}, vol. 211, p.
  104626, 2020. [Online]. Available:
  \url{https://www.sciencedirect.com/science/article/pii/S0045793020301985}
\BIBentrySTDinterwordspacing

\bibitem{Stabile2017CAIM}
G.~Stabile, S.~Hijazi, A.~Mola, S.~Lorenzi, and G.~Rozza, ``{POD-Galerkin
  reduced order methods for CFD using Finite Volume Discretisation: vortex
  shedding around a circular cylinder},'' \emph{Communications in Applied and
  Industrial Mathematics}, vol.~8, no.~1, pp. 210--236, (2017).

\bibitem{Stabile2017CAF}
\BIBentryALTinterwordspacing
G.~Stabile and G.~Rozza, ``{Finite volume POD-Galerkin stabilised reduced order
  methods for the parametrised incompressible Navier-Stokes equations},''
  \emph{Computers \& Fluids}, 2018. [Online]. Available:
  \url{https://www.sciencedirect.com/science/article/pii/S0045793018300422}
\BIBentrySTDinterwordspacing

\bibitem{Kochkove2101784118}
\BIBentryALTinterwordspacing
D.~Kochkov, J.~A. Smith, A.~Alieva, Q.~Wang, M.~P. Brenner, and S.~Hoyer,
  ``Machine learning{\textendash}accelerated computational fluid dynamics,''
  \emph{Proceedings of the National Academy of Sciences}, vol. 118, no.~21,
  2021. [Online]. Available:
  \url{https://www.pnas.org/content/118/21/e2101784118}
\BIBentrySTDinterwordspacing

\bibitem{AccelerateCFD}
{Illinois Rocstar LLC's}, ``{AcclerateCFD Community Edition},''
  \url{https://github.com/IllinoisRocstar/AccelerateCFD_CE}.

\bibitem{ITHACA-FV}
{SISSA mathLab}, ``{ITHACA-FV},'' https://mathlab.sissa.it/ITHACA-FV.

\bibitem{AndreWeiner}
A.~Weiner, ``{Machine learning applied to CFD },''
  \url{https://github.com/AndreWeiner/machine-learning-applied-to-cfd}.

\bibitem{PranshuPant}
P.~Pant, ``{DL-ROM : Deep Learning for Reduced Order Modelling},''
  \url{https://github.com/pranshupant/DL-ROM}.

\bibitem{PythonFOAM}
R.~Maulik, ``{In-situ data analyses and machine learning with OpenFOAM and
  Python},'' \url{https://github.com/argonne-lcf/PythonFOAM}.

\bibitem{zwart_ecological_2020}
\BIBentryALTinterwordspacing
S.~P. Zwart, ``The {Ecological} {Impact} of {High}-performance {Computing} in
  {Astrophysics},'' \emph{Nature Astronomy}, vol.~4, no.~9, pp. 819--822, Sep.
  2020, arXiv: 2009.11295. [Online]. Available:
  \url{http://arxiv.org/abs/2009.11295}
\BIBentrySTDinterwordspacing

\bibitem{freitag2021real}
C.~Freitag, M.~Berners-Lee, K.~Widdicks, B.~Knowles, G.~S. Blair, and
  A.~Friday, ``{The real climate and transformative impact of ICT: A critique
  of estimates, trends, and regulations},'' \emph{Patterns}, vol.~2, no.~9, p.
  100340, 2021.

\bibitem{strubell_energy_2019}
\BIBentryALTinterwordspacing
E.~Strubell, A.~Ganesh, and A.~McCallum, ``Energy and {Policy} {Considerations}
  for {Deep} {Learning} in {NLP},'' \emph{arXiv:1906.02243 [cs]}, Jun. 2019,
  arXiv: 1906.02243. [Online]. Available: \url{http://arxiv.org/abs/1906.02243}
\BIBentrySTDinterwordspacing

\bibitem{kitwarewebsite}
\BIBentryALTinterwordspacing
{Kitware Inc.} Building solutions on open source technologies. [Online].
  Available: \url{https://www.kitware.com/open-source/}
\BIBentrySTDinterwordspacing

\bibitem{vtkjsWeb}
------. {vtk.js}. \url{https://kitware.github.io/vtk-js}.

\bibitem{paraviewWeb}
------. {ParaViewWeb}. \url{https://www.paraview.org/web}.

\bibitem{paraviewGlanceWeb}
------. {ParaView Glance}. \url{https://kitware.github.io/paraview-glance}.

\bibitem{trame}
------. {trame}. \url{https://kitware.github.io/trame/ }.

\bibitem{HesthavenRozzaStamm2015}
\BIBentryALTinterwordspacing
J.~S. Hesthaven, G.~Rozza, and B.~Stamm, \emph{{Certified Reduced Basis Methods
  for Parametrized Partial Differential Equations}}, ser. SpringerBriefs in
  Mathematics.\hskip 1em plus 0.5em minus 0.4em\relax Springer International
  Publishing, 2015. [Online]. Available:
  \url{https://link.springer.com/book/10.1007/978-3-319-22470-1}
\BIBentrySTDinterwordspacing

\bibitem{RBniCS}
{RBniCS Project}. {RBniCS - reduced order modelling in FEniCS}.
  \url{https://www.rbnicsproject.org}.

\bibitem{ARGOS}
\BIBentryALTinterwordspacing
{SISSA mathLab}. Advanced reduced groupware online simulation platform.
  [Online]. Available: \url{https://argos.sissa.it}
\BIBentrySTDinterwordspacing

\bibitem{kitwareDeepLearning}
F.~Mazen and A.~Schieb, ``{Deep-learning surrogate models in ParaView: Viewing
  inference results and monitoring the training process in real time with
  Catalyst},''
  \url{https://www.kitware.com/deep-learning-surrogate-models-in-paraview-viewing-inference-results-and-monitoring-the-training-process-in-real-time-with-catalyst}.

\bibitem{rom.js}
{SIMZERO}. {rom.js}. \url{https://github.com/simzero-oss/rom-js}.

\bibitem{cfd.xyz}
------. {cfd.xyz viewer}. \url{https://github.com/simzero-oss/cfd-xyz-viewer}.

\bibitem{eigenweb}
G.~Guennebaud, B.~Jacob \emph{et~al.}, ``Eigen v3,''
  http://eigen.tuxfamily.org, 2010.

\bibitem{splinter}
B.~Grimstad \emph{et~al.}, ``{SPLINTER: a library for multivariate function
  approximation with splines},'' \url{http://github.com/bgrimstad/splinter},
  2015, accessed: 2022-03-19.

\bibitem{vtk}
W.~Schroeder, K.~Martin, and B.~Lorensen, \emph{{The Visualization Toolkit - An
  Object-Oriented Approach To 3D Graphics}}, 4th~ed.\hskip 1em plus 0.5em minus
  0.4em\relax Kitware, Inc., 2006.

\bibitem{salmoiraghi2016advances}
F.~Salmoiraghi, F.~Ballarin, G.~Corsi, A.~Mola, M.~Tezzele, G.~Rozza
  \emph{et~al.}, ``Advances in geometrical parametrization and reduced order
  models and methods for computational fluid dynamics problems in applied
  sciences and engineering: overview and perspectives,'' in \emph{VII European
  Congress on Computational Methods in Applied Sciences and Engineering},
  vol.~1.\hskip 1em plus 0.5em minus 0.4em\relax Institute of Structural
  Analysis and Antiseismic Research School of Civil~…, 2016, pp. 1013--1031.

\bibitem{RossbergWebAssemblyCoreSpecification}
\BIBentryALTinterwordspacing
A.~Rossberg, ``{{WebAssembly Core Specification}},'' 2019. [Online]. Available:
  \url{https://www.w3.org/TR/wasm-core-1/}
\BIBentrySTDinterwordspacing

\bibitem{hijazi_data-driven_2020}
\BIBentryALTinterwordspacing
S.~Hijazi, G.~Stabile, A.~Mola, and G.~Rozza,
  ``\BIBforeignlanguage{en}{Data-driven {POD}-{Galerkin} reduced order model
  for turbulent flows},'' \emph{\BIBforeignlanguage{en}{Journal of
  Computational Physics}}, vol. 416, p. 109513, Sep. 2020. [Online]. Available:
  \url{https://www.sciencedirect.com/science/article/pii/S0021999120302874}
\BIBentrySTDinterwordspacing

\end{thebibliography}
